\documentclass[preprint2]{aastex63}

\shorttitle{Dayside land}
\shortauthors{Macdonald et al.}

\graphicspath{{./}{figures/}}

\begin{document}

\title{Climate uncertainties caused by unknown land distribution on habitable M-Earths}

\correspondingauthor{Evelyn Macdonald}
\email{evelyn.macdonald@mail.utoronto.ca}

\author[0000-0001-5540-3817]{Evelyn Macdonald}
\affiliation{Department of Physics, University of Toronto, Toronto, ON, Canada}

\author[0000-0001-6774-7430]{Adiv Paradise}
\affiliation{David A. Dunlap Department of Astronomy and Astrophysics, University of Toronto, ON, Canada}

\author{Kristen Menou}
\affiliation{Department of Physical and Environmental Sciences, University of Toronto,
Scarborough, ON, Canada}
\affiliation{David A. Dunlap Department of Astronomy and Astrophysics, University of Toronto, ON, Canada}
\affiliation{Department of Physics, University of Toronto, Toronto, ON, Canada}

\author{Christopher Lee}
\affiliation{Department of Physics, University of Toronto, Toronto, ON, Canada}

\begin{abstract}
A planet's surface conditions can significantly impact its climate and habitability. In this study, we use the 3D general circulation model ExoPlaSim to systematically vary dayside land cover on a synchronously rotating, temperate rocky planet under two extreme and opposite continent configurations, in which either all of the land or all of the ocean is centred at the substellar point. We identify water vapour and sea ice as competing drivers of climate, and we isolate land-dependent regimes under which one or the other dominates. We find that the amount and configuration of land can change the planet's globally averaged surface temperature by up to $\sim$20~K, and its atmospheric water vapour content by several orders of magnitude. The most discrepant models have partial dayside land cover with opposite continent configurations. Since transit spectroscopy may permit observations of M-dwarf planets' atmospheres, but not their surfaces, these land-related climate differences likely represent a limiting uncertainty in a given planet's climate, even if its atmospheric composition is known. Our results are robust to variations in atmospheric CO$_2$ concentration, stellar temperature, and instellation.
\end{abstract}


\section{Introduction} \label{sec:intro}

The growing number of known exoplanets has led to extensive research on which types of planets could be habitable. M-Earths, defined here as Earth-sized planets with M-dwarf host stars, are possible targets for characterization with next-generation instruments like the James Webb Space Telescope. Several such planets have been discovered in recent years (e.g., \citealt{Anglada-Escude2016, Dittmann2017, Gillon2017, Gilbert2020}), but their climates have yet to be characterised observationally.

M-Earths are likely to differ drastically from Earth, most notably due to synchronous rotation, proximity to their host stars, and differences in stellar temperature. Recent work has shown that both atmospheric and surface properties have a significant impact on an M-Earth's climate and habitability. Studies have analyzed the general circulation of synchronously rotating exoplanets using 3D simulations (e.g., \citealt{Joshi1997, Merlis2010, Turbet2016, Boutle2017, Turbet2018, Komacek2019, Yang2019, Fauchez2021, Sergeev2021, Turbet2021}). Others have examined the effects of surface composition and albedo on M-Earth climates \citep{Shields2013, Shields2016, Shields2018, Rushby2019, Rushby2020}.  

Temperate, synchronously rotating M-Earths are expected to have ``eyeball" climates \citep{Pierrehumbert2011}, meaning that the planet is mostly frozen except for a deglaciated circular region around the substellar point, where the irradiation is highest. Other studies \citep{Lewis2018, Salazar2020} have found that large continents in the deglaciated region significantly impact an M-Earth's climate. 

In this study, we systematically vary dayside land fraction and configuration on an M-Earth using a large set of 3D climate simulations. We find that land has little effect on the general circulation, but can cause large changes in temperature and humidity. Our work highlights the need for detailed surface conditions in M-Earth climate models. All of our simulation outputs, as well as the files needed to reproduce them, are available in a permanent Dataverse repository\footnote{\url{https://doi.org/10.5683/SP3/JCLM0O}} \citep{Paradise2020, Macdonald2021}.

We describe our model in section \ref{sec:model}, present our results in section \ref{sec:results}, and discuss our findings in section \ref{sec:discussion}.

\section{Model Setup} \label{sec:model}

\subsection{Description of the GCM}

We use ExoPlaSim \citep{Paradise2021tlcoords,Paradise2021}, a version of the intermediate-complexity spectral 3D general circulation model (GCM) PlaSim \citep{Fraedrich2005} adapted for exoplanet simulations. The model has a slab ocean with fixed mixed layer depth and infinite water supply, such that water entering and leaving the ocean is not conserved. Land is modelled as a water bucket with runoff and evaporation. We do not consider topography.

For most of our simulations, we use a resolution of T63, or 96 latitudes by 192 longitudes, with 20 vertical levels. We have found this resolution to be the best balance of speed and precision, with each simulated year taking about 120 minutes of computing time on 32 cores on a modern workstation. We also include some simulations at T42, or 64 latitudes by 128 longitudes. At lower resolutions, the nightside is artificially cloudy and moist due to Gibbs oscillations in the spectral transform, which arise due to day-night near-discontinuities in spectral GCMs (e.g., \citealt{Navarra1994}). The Gibbs oscillations manifest as rings of alternating amplitude in clouds, precipitation, water vapour, and other fields. They are most visible on the nightside, particularly in cold and dry climates. We damp the Gibbs oscillations using an exponential filter in the transforms from gridpoint to spectral space and back; see \citet{Paradise2021} for a description of the use of filters in ExoPlaSim. In our filtered models, the clouds become more structured, particularly on the nightside, suggesting that the filters reveal real physical effects that had been obscured by the Gibbs oscillations.

\subsection{Simulation initialization}

We base our simulations on the planet Proxima Centauri B \citep{Anglada-Escude2016}. \citet{Turbet2016, Boutle2017, DelGenio2019} considered possible habitable scenarios for this planet, while \citet{Lewis2018, Salazar2020} showed that its climate is sensitive to its dayside land fraction. In this study, we assume the planet to be synchronously rotating on a circular orbit. Unless otherwise stated, each simulation has the parameters listed in Table \ref{tab:params}. We assume a baseline of a 1~bar N$_2$ atmosphere plus 1~millibar of CO$_2$, except in simulations in which we vary the pCO$_2$. We use ExoPlaSim’s default land albedo model, which yields an average value of $\sim$0.22 based on a mix of surface types. We allow ExoPlaSim to calculate the ocean, ice, snow, and cloud albedos in its shortwave and longwave bands based on a stellar blackbody spectrum of 3000~K, as described in \citet{Paradise2021}. The albedos for ocean and ice are around 0.04 and 0.41, respectively.

\begin{table}[h!]
    \centering
    \begin{tabular}{c|c}
        Parameter &  Value \\
        \hline
        Semimajor axis & 0.0045 AU \\
        Instellation & 881.7 W/m$^2$ \\
        Orbital period & 11.86 days \\
        Radius & 1.12 R$_\oplus$ \\
        Gravity & 10.9 m/s$^2$ \\
        Stellar temperature & 3000 K
    \end{tabular}
    \caption{Parameters for Proxima Centauri B.}
    \label{tab:params}
\end{table}

\begin{figure}[h!]
\centering
\includegraphics[width=\columnwidth]{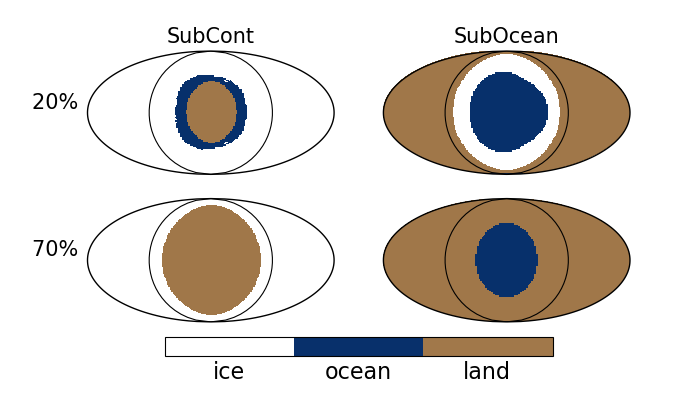}
\caption{Maps of the two main land regimes considered: substellar continent (SubCont, left) and substellar ocean (SubOcean, right), with 20\% (top) and 70\% (bottom) dayside land cover. These maps are Mollweide projections centred at the substellar point, with the black circle representing the terminator.}
\label{fig:diagram}
\end{figure}

We consider two end-member landmap classes (figure \ref{fig:diagram}): substellar continent (SubCont), featuring a circular continent centred at the substellar point, somewhat similar to \citet{Lewis2018,Salazar2020}; and substellar ocean (SubOcean), in which a circular ocean is centred at the substellar point with land covering the nightside and remainder of the dayside. For each landmap class, we vary the dayside land fraction from 0 to 100\% in increments of 10\%, keeping the land distribution symmetric around the substellar point. While our landmap classes are intended as extreme cases, they may accurately represent some real planets, as \citet{Leconte2018} suggests that land on a synchronously rotating planet may preferentially be at the substellar or antistellar point. Using ExoPlaSim's random continent generation module, we also include models with randomly distributed dayside continents (RandCont). We run 10-12 such models at each of 20\%, 40\%, 60\%, and 80\% dayside land cover, at a resolution of T42.

\section{Results} \label{sec:results}

\subsection{General circulation} \label{sec:gc}

\begin{figure}[h!]
\centering
\includegraphics[width=\columnwidth]{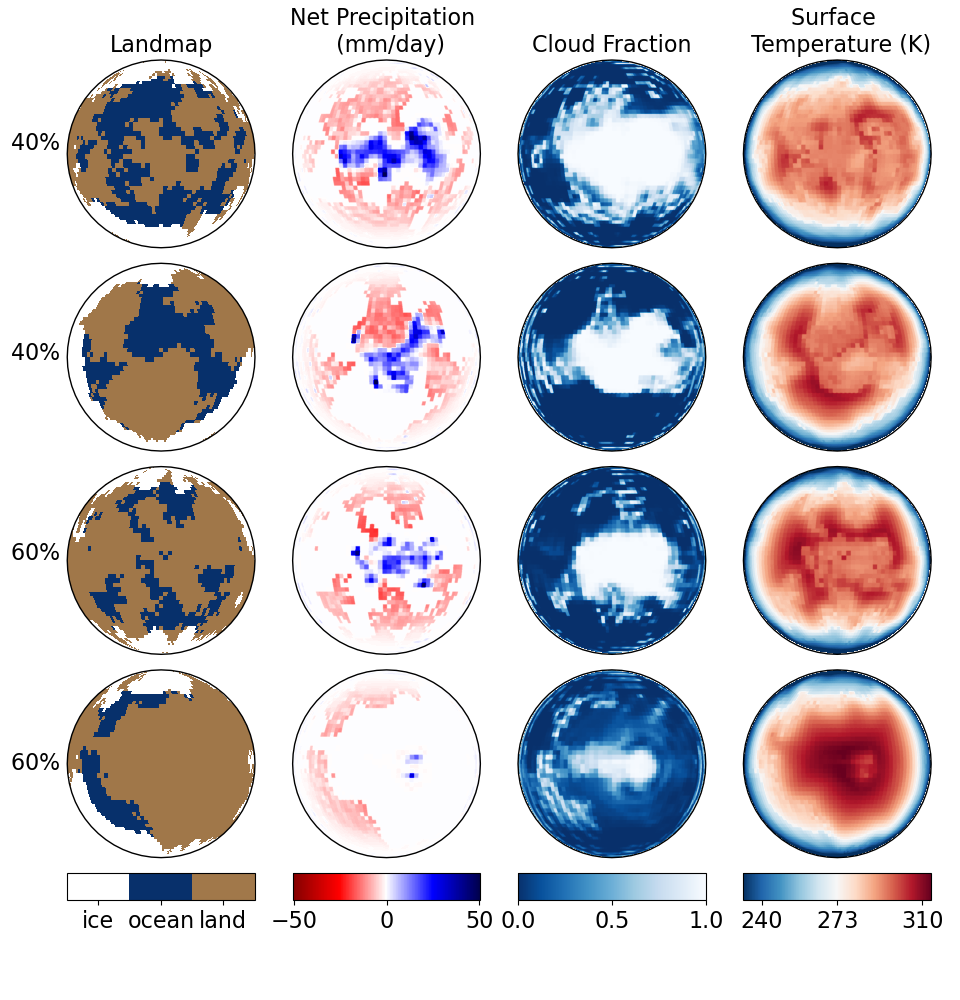}
\caption{Dayside orthographic projections of sample RandCont models with 40-60\% dayside land cover (row labels). Left to right: landmap, net precipitation (defined as precipitation minus evaporation), cloud fraction, and surface temperature. Clouds and precipitation are always concentrated near the substellar point, but their exact distribution depends on the shape of the land. Evaporation takes place over ocean elsewhere on the dayside. Surface temperatures are highest on dry land.} 
\label{fig:rcmaps}
\end{figure}

Our simulations have the general circulation regime of Rhines rotators \citep{Haqq-Misra2018}. There is one main circulation cell from the substellar point to the nightside. Warm air rises in the substellar region, releasing its moisture, and travels outward, preferentially eastward. The air then descends near the terminator and returns to the substellar region along the surface; if there is ice-free ocean on the outer dayside, the air accumulates moisture during its return. The substellar region is cloudy in all land configurations; elsewhere on the dayside, clouds only form over ocean, and land is hot and dry. Figure \ref{fig:rcmaps} shows a selection of RandCont climates with 40\% dayside land cover. Each of these models has a rainy and cloudy substellar region whose shape depends on the configuration of the continents. Although some moisture also travels to the nightside and can become trapped as ice and snow, this circulation regime can still maintain a substellar hydrological cycle, as noted by \citet{Ding2020}. Nightside ice trapping does not affect dayside water inventories in our models because ExoPlaSim's oceans have an infinite water supply. 

\begin{figure}[h!]
    \centering
    \includegraphics[width=\columnwidth]{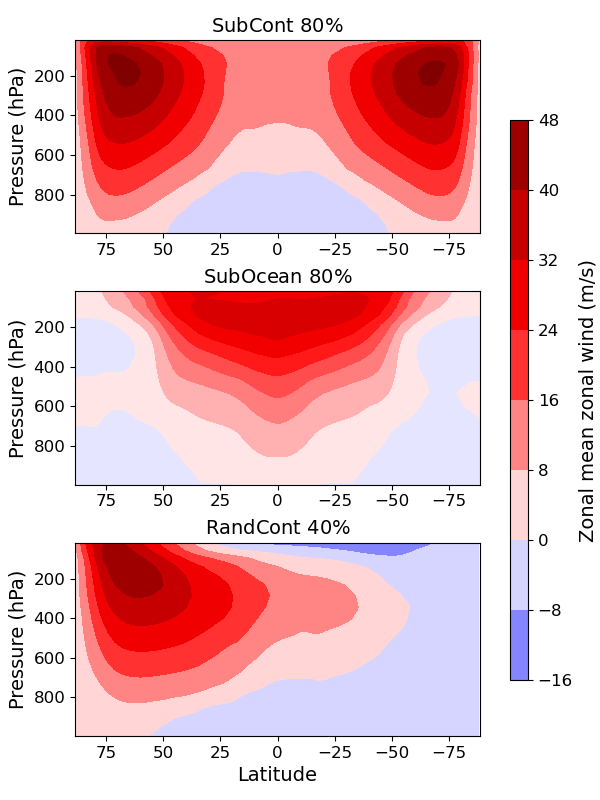}
    \caption{Zonal mean zonal winds for sample models of each landmap type. Titles indicate the dayside land cover percentage. With the exception of a few RandCont models and high-land-fraction SubOcean models, all of our simulations have two mid-latitude zonal jets, whose strength and shape depend weakly on the landmap.}
    \label{fig:zmz}
\end{figure}

As is typical of Rhines rotators, most of our simulations also have two mid-latitude eastward jets, whose strength and shape depend on the landmap. Some SubOcean models with high dayside land fraction have a single high-altitude eastward jet and two lower, slower, high-latitude counterrotating currents, somewhat similar to the E4 circulation profile of \citet{Lewis2018}. Figure \ref{fig:zmz} shows the zonal mean zonal wind for a sample simulation of each landmap type. 

\begin{figure}
    \centering
    \includegraphics[width=\columnwidth]{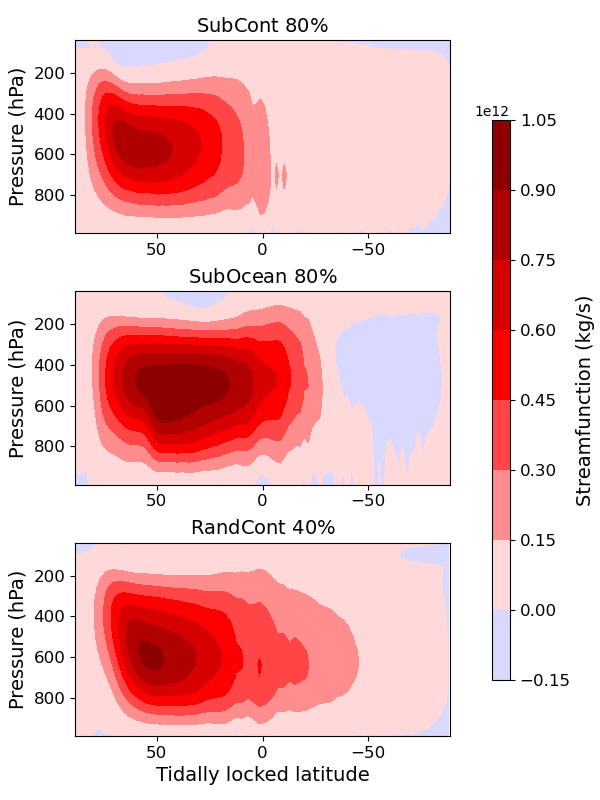}
    \caption{Tidally locked streamfunctions for the same sample models as figure \ref{fig:zmz}, one of each landmap type. The terminator is at a tidally locked latitude of 0$^\circ$, with positive tidally locked latitudes representing the dayside, and negative ones, the nightside. Air circulates from the substellar point to the nightside in one large cell whose strength depends on the landmap.}
    \label{fig:sf}
\end{figure}

We illustrate the day-night overturning circulation for these sample simulations in figure \ref{fig:sf} using the tidally locked streamfunction \citep{Hammond2021, Paradise2021}. This streamfunction is computed in a tidally locked coordinate system (\citealt{Paradise2021}, based on \citealt{Koll2015,Hammond2021}), in which the equator represents the terminator and the substellar and antistellar points are at $90^\circ$ and $-90^\circ$, respectively.  The tidally locked streamfunctions show air rising near the substellar point, travelling toward the terminator and to the nightside, and then returning to the substellar region along the dayside surface. The exact shape and strength of the streamfunction depends on the landmap. 

Despite their similarities in general circulation, our simulations have a large range of climates due to the effect of land configuration on water availability. The amount of ice-free ocean on the dayside, which depends on both the iceline location and the amount of land in the deglaciated region, determines how much water vapour a planet can have in its atmosphere. Since SubCont landmaps have all of their land in the warmest part of the planet, those with high land fraction have little to no ice-free ocean, so their climates are dry. On the other hand, SubOcean models have all of their ocean in the warmest part of the planet, so they always have a significant amount of water vapour in their atmospheres.

\begin{figure*}
\centering
\includegraphics[width=\textwidth]{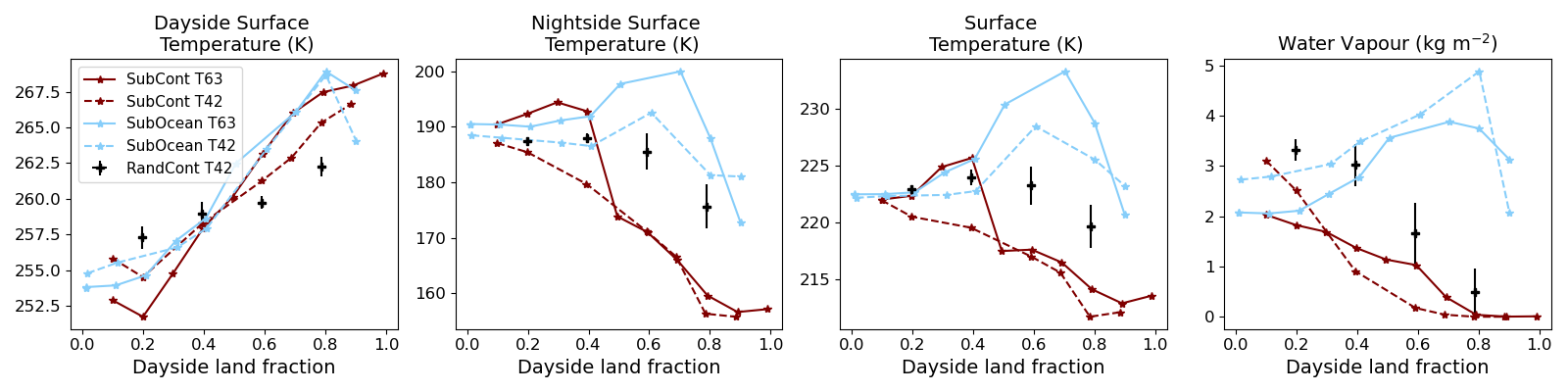}
\caption{Dayside, nightside, and globally averaged surface temperature, and globally averaged atmospheric water vapour, as a function of dayside land fraction, for our three classes of landmap. SubCont and SubOcean trends are shown at both T63 and T42; there is relatively good agreement between the two resolutions. Dayside surface temperature increases as a function of land fraction for all landmap classes; nightside temperature always decreases, but at different rates, indicating that the models recirculate heat to the nightside with different efficiencies. The globally averaged temperature and water vapour trends differ significantly between SubCont and SubOcean models, and are most discrepant at partial dayside land cover. These two landmap classes are extrema by which the RandCont models are mostly bounded. We attribute these climate trends to differences in moisture availability and surface albedo; see the main text for details.}
\label{fig:trends}
\end{figure*}

Figure \ref{fig:trends} shows the average surface temperature and atmospheric water vapour content as a function of dayside land fraction for our three landmap classes. We also include T42 trends for SubCont and SubOcean climates to facilitate comparison with the RandCont models. The RandCont models generally fall between the SubOcean and SubCont models, which represent extreme and opposite scenarios. The globally averaged surface temperature depends on the water vapour content, which is highest for SubOcean models with partial dayside land cover because these have the most ice-free ocean, as discussed in section \ref{sec:so}. The RandCont models also highlight the more universal trends of higher dayside temperatures, cool nightsides, and lower overall temperatures at high dayside land fraction; however, there is considerable spread between individual RandCont models at a given land fraction because these models have significant differences in spatially resolved surface temperature, wind speed, precipitation, and cloud cover, as shown in figure \ref{fig:rcmaps}. SubCont and SubOcean climates are discussed in detail below.

\subsection{SubCont climates} \label{sec:sc}

\begin{figure}[h!]
\centering
\includegraphics[width=\columnwidth]{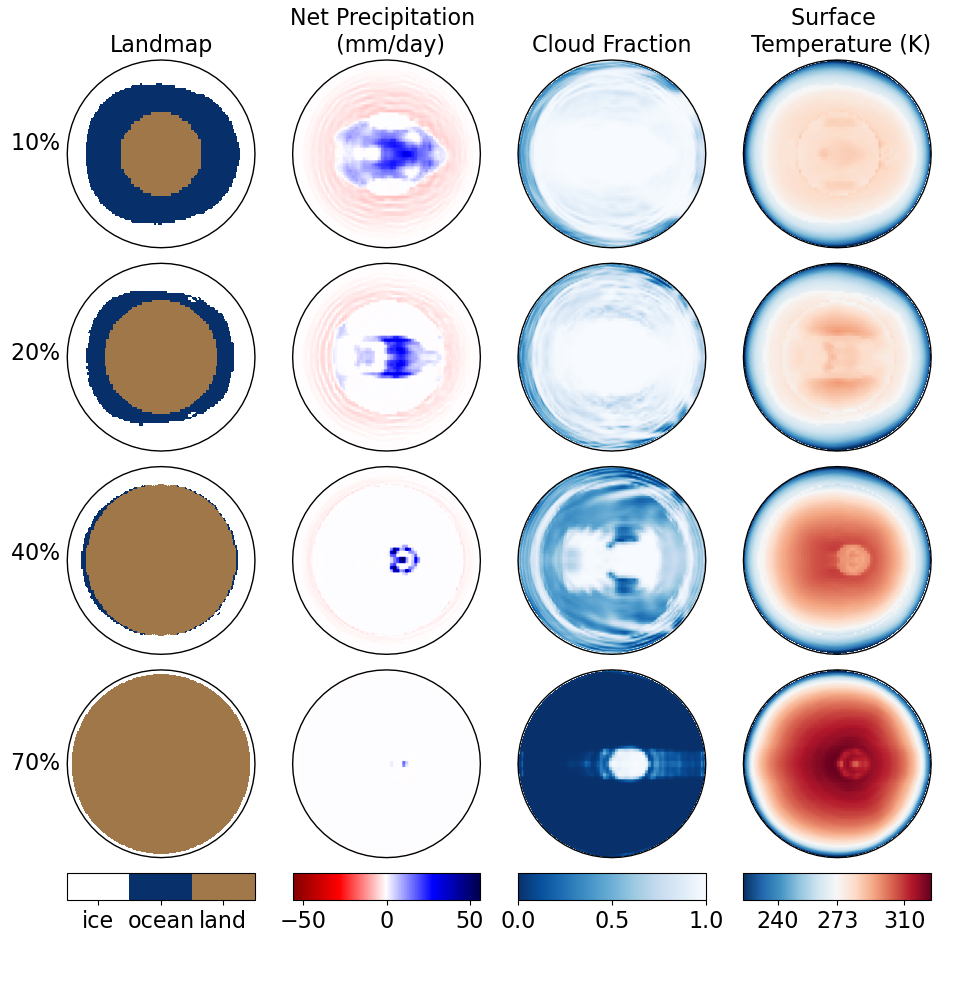}
\caption{Dayside maps of climate variables for SubCont models with a range of dayside land fractions (row labels). Left to right: landmap, net precipitation, cloud fraction, and surface temperature. Models with high land fraction have hotter and drier daysides, with dayside clouds and precipitation mostly confined to a small substellar region.}
\label{fig:scmaps}
\end{figure}

The configuration of the SubCont landmaps makes moisture availability particularly sensitive to land fraction. The smallest substellar continents have little effect on atmospheric water vapour, since most evaporation takes place away from the substellar point. However, water vapour drops significantly once the continent edge reaches the iceline, at which point the average temperature also decreases. The little water that evaporates from sea ice continues to precipitate out over the substellar region, although the precipitation region is both drier and smaller than in models with ice-free ocean. Dayside maps of SubCont climates at a range of land fractions are shown in figure \ref{fig:scmaps}.

\begin{figure*}
    \centering
    \includegraphics[width=\textwidth]{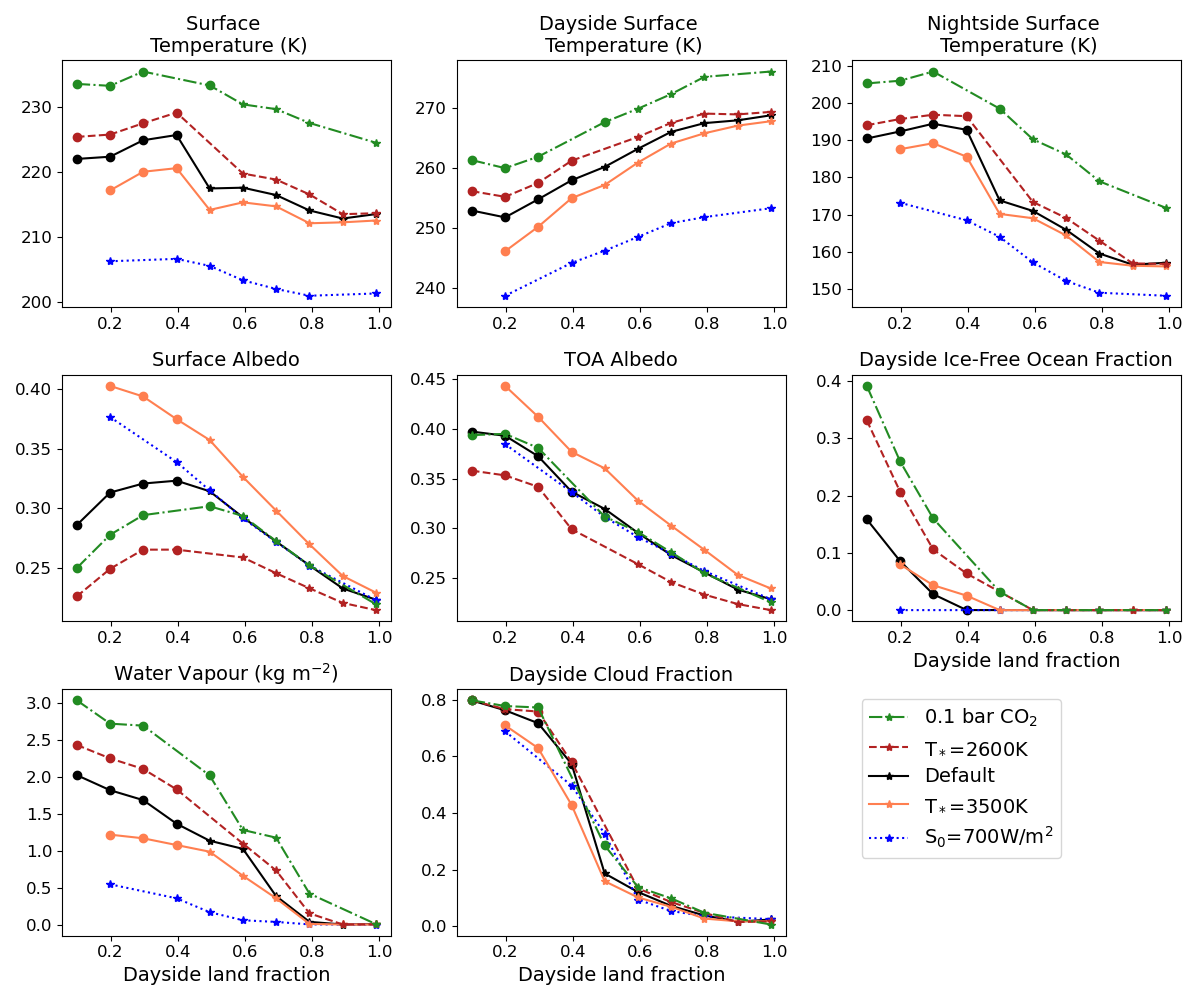}
    \caption{Comparison of SubCont models with high pCO$_2$ (green), stellar temperatures of 3500~K (orange) and 2600~K (red), and low instellation (blue). Circles are added to the curve to point out models that have some ice-free ocean. In all cases, as the continent size increases, the dayside surface temperature increases, while the nightside and globally averaged surface temperatures and atmospheric water vapour decrease. Albedos are highest at low land fraction, where there are the most ice and clouds. Changing the pCO$_2$ shifts the entire temperature curves vertically without significant changes to their shapes. The colder models only have ice-free ocean at low land fraction. Decreasing the stellar temperature but not the instellation increases the planet's temperature because ice and cloud albedos are lower at longer wavelengths; this difference has the strongest impact on low-land-fraction models because they have more sea ice. Conversely, raising the stellar temperature makes ice and clouds brighter, so these low-land-fraction models are cooler for hotter stars.}
    \label{fig:SCtrends}
\end{figure*}

Figure \ref{fig:SCtrends} shows the landmap-dependent climate trends of SubCont models in response to varying CO$_2$ concentration, stellar temperature, and instellation. We show temperatures, surface and top-of-atmosphere (TOA) albedos, fraction of the dayside that is ice-free ocean, atmospheric water vapour, and cloud cover. Models that have sea ice present are marked by circles. All averaged albedos in this and other figures are for the dayside only, and are weighted by cross-sectional area from the substellar point to represent the actual fraction of reflected incoming stellar radiation. Water vapour and globally averaged surface temperature generally drop once the dayside ice-free ocean fraction reaches zero. The surface albedo increases with land fraction until this point because land is brighter than ocean; once the continent reaches the iceline, increasing the land fraction replaces sea ice with lower-albedo land, and the surface albedo decreases again. The TOA albedo is dominated by clouds and so decreases steadily with increasing land fraction as the substellar precipitation region shrinks.

The overall climate trends seen in figure \ref{fig:trends} are preserved across the perturbations. More CO$_2$ in the atmosphere makes the planet hotter and rainier overall, but has little effect on the land dependence of the climate trends. Conversely, decreasing the instellation results in a colder, drier planet whose ocean is entirely covered in sea ice. Changing the stellar temperature but not the instellation affects the temperature because ice and clouds have lower albedos at longer wavelengths. At low land fraction, where there is the most ice, the coolest host star corresponds to the warmest planet. Water vapour is also more opaque in the infrared, so planets with cooler host stars absorb more radiation directly in their atmospheres \citep{Shields2013}. As instellation, stellar temperature, and rotation rate are related for synchronously rotating habitable-zone planets, our modifications of these parameters are meant only to illustrate how the climates respond, without attempting to model realistic systems. 

\subsection{SubOcean climates} \label{sec:so}

\begin{figure}[h!]
\centering
\includegraphics[width=\columnwidth]{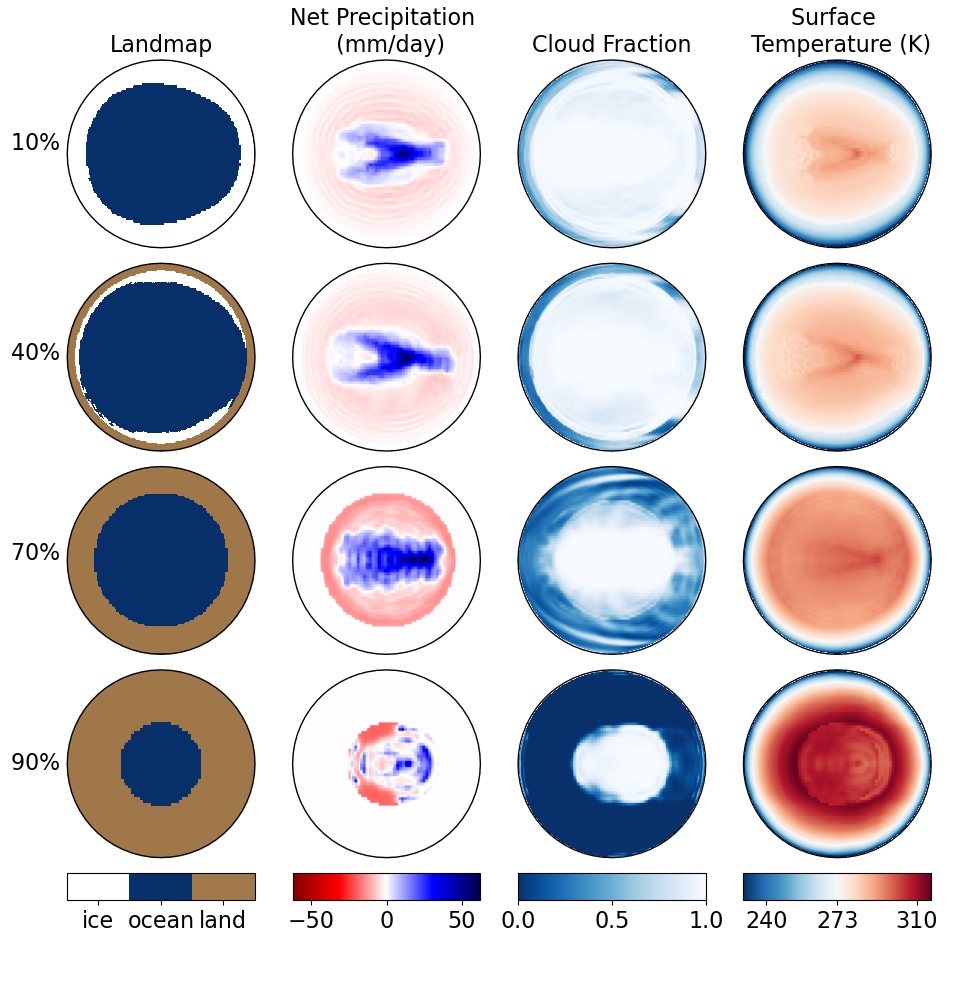}
\caption{Dayside maps of climate variables for SubOcean models with a range of dayside land fractions (row labels). Left to right: landmap, net precipitation, cloud fraction, and surface temperature. As with the SubCont models, the cloud and precipitation coverage depends heavily on the land fraction, and models with little ice-free ocean have hot, dry land.}
\label{fig:somaps}
\end{figure}

Figure \ref{fig:somaps} shows dayside maps of several SubOcean models. As with SubCont models, daysides are hot and dry with limited cloud and precipitation cover in models with little ice-free ocean.

Unlike SubCont models, all SubOcean models have some ice-free ocean, but those with high land fraction do not have any sea ice. Again, the configuration of land and sea ice determines the size of the ice-free ocean, which influences the surface temperature through its effect on atmospheric water vapour. \citet{Checlair2017, Checlair2019} used both an energy balance model and idealized GCM simulations to show that the amount of ice on a tidally locked aquaplanet for a given instellation can be predicted based on assumed ocean and ice albedos. We find that the iceline's location is also heavily influenced by the presence of land, because the latter changes the planet's albedo. 

\begin{figure}[h!]
    \centering
    \includegraphics[width=\columnwidth]{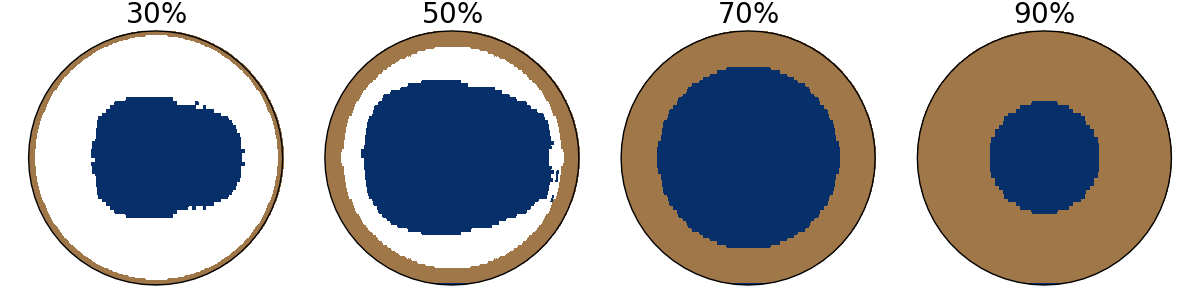}
    \caption{The ice albedo feedback in SubOcean models. Dayside land cover (panel titles) increases to the right. In models that have sea ice, the ice-free ocean decreases in size as the land fraction decreases due to a positive albedo feedback whereby dayside sea ice cools the planet, causing the formation of more ice. The location of the iceline in an eyeball climate is therefore heavily dependent on surface conditions.}
    \label{fig:icealbedo}
\end{figure}

The right branch of the SubOcean global temperature and water vapour curves in figure \ref{fig:trends} shows a tendency for planets to be cooler and drier at high land fraction, where the substellar ocean is small. However, both temperature and water vapour peak at partial dayside land cover; on the left branch of these curves, climates become colder and drier again as land fraction decreases, despite the increasing substellar ocean size. We attribute this shape to an albedo feedback, illustrated in figure \ref{fig:icealbedo}. The albedos in these models are such that $\alpha_\mathrm{ocean} < \alpha_\mathrm{land} < \alpha_\mathrm{ice}$; therefore, decreasing the land fraction replaces land with lower-albedo water, making the whole planet darker. However, once the ocean is large enough for sea ice to form around its edges, further decreasing the land fraction replaces land with high-albedo sea ice instead, so the planet cools and the iceline moves toward the substellar point, which increases the overall albedo even more. Consequently, further increasing the ocean size decreases the amount of ice-free ocean. This feedback would be reversed in the case where land is brighter than ice, which is possible with a cool host star and high-albedo land. Since most precipitation falls near the substellar point, SubOcean models have little snow on land, so the surface albedo of cold continents is close to that of temperate land. 

\begin{figure*}
    \centering
    \includegraphics[width=\textwidth]{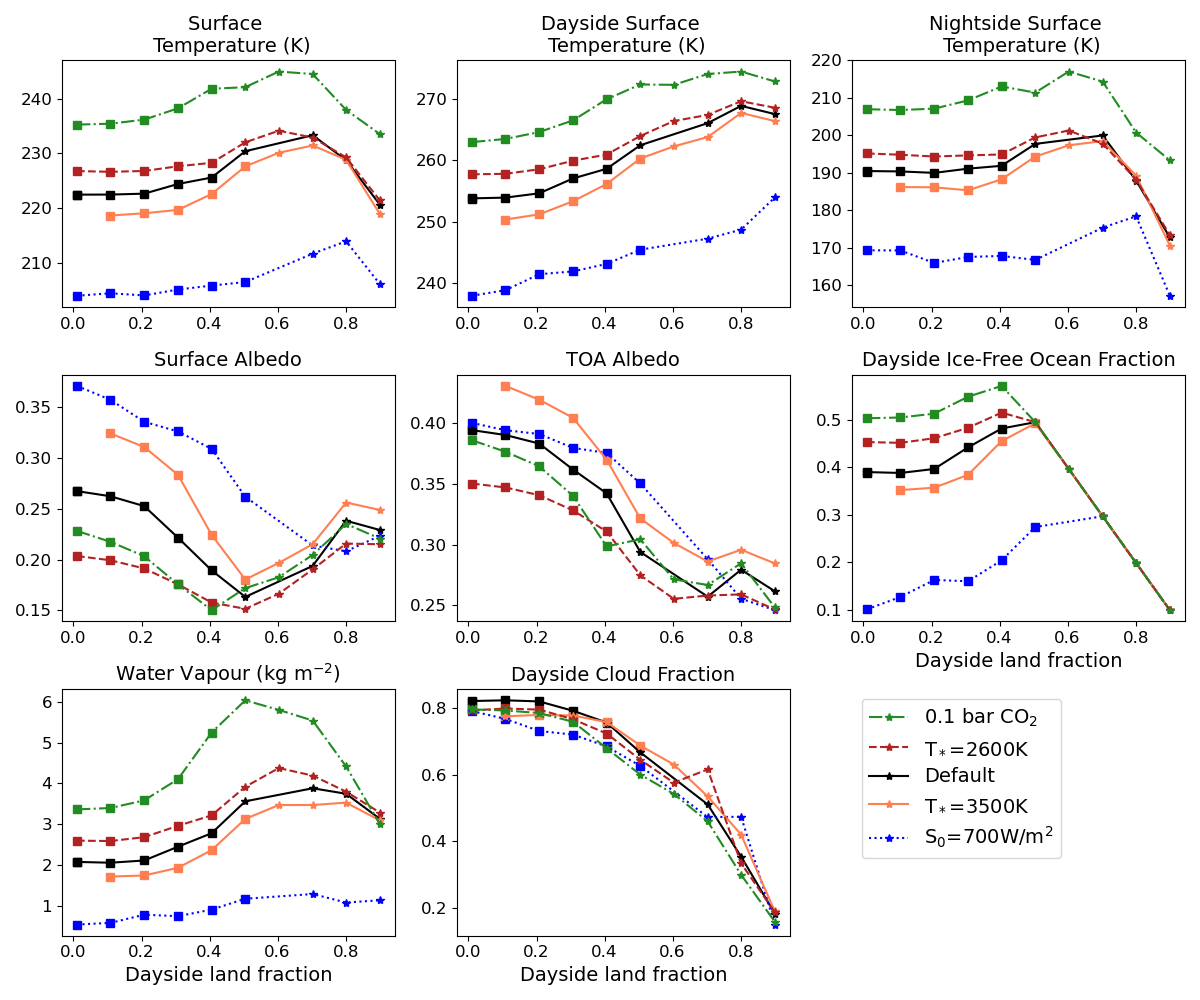}
    \caption{Similar to figure \ref{fig:SCtrends} but for SubOcean models. Models with non-zero sea ice are marked with squares. Globally averaged temperature and water vapour always peak around the highest-land-fraction model without sea ice, where the ice-free ocean is largest. Increasing the pCO$_2$ or decreasing the instellation shifts the entire temperature curves vertically. The ice-free ocean fraction is much lower, and the surface albedo is much higher, in the low-instellation models. The stellar temperature also has a significant effect on the surface albedo and temperature because ice albedo is higher at shorter wavelengths; consequently, SubOcean landmaps have stronger ice albedo feedbacks, which generate more climate uncertainty, when the host star is hotter. The albedo differences between models are most pronounced at low land fraction, where there is the most ice.}
    \label{fig:SOtrends}
\end{figure*}

We show the climate trends of SubOcean models in response to perturbations in CO$_2$ concentration, stellar temperature, and instellation in figure \ref{fig:SOtrends}. Models containing sea ice are indicated by squares. The surface temperature and water vapour trends generally peak around the highest-land-fraction model that contains some sea ice; this is where the ice-free ocean area is maximized, the surface albedo is minimized, and the planet is warmest. As in the SubCont models, the effect of stellar temperature on ice albedo and planet temperature is strongest at low land fraction. In particular, the ice albedo feedback is more pronounced at a higher stellar temperature due to the ice's higher albedo. This effect results in a larger land-induced climate uncertainty for warmer host stars.

\section{Discussion} \label{sec:discussion}

We have established that dayside land on a synchronously rotating M-Earth has a significant impact on its climate. Although we observe only minor changes in the general circulation, land influences the climate through its effects on surface albedo and availability of water for evaporation. Models with substellar continents become cooler and drier as we replace ocean with higher-albedo land; they are able to maintain a substellar hydrological cycle, but most of the land is hot and dry when the continent is large. On the other hand, substellar ocean climates are warmest and the most humid at partial dayside land cover, where the ice-free ocean is largest. These models show competition between the opposite effects of ice albedo and atmospheric water vapour concentration, with the former dominating at low land fraction and the latter at high land fraction. These climate trends are robust to changes in CO$_2$ concentration, instellation, and stellar spectrum, even as they shift the location of the iceline. Models with randomly generated continents have less variability in averaged climate trends, but more spatially resolved differences, than the other landmap classes. The general trend of warming daysides and increasing day-night temperature contrasts as the land fraction increases is present for all landmap classes.

Our observed SubCont climate trends differ from those of \citet{Lewis2018, Salazar2020}, which also differ from each other. We attribute this discrepancy to a combination of land albedo differences and the fact that these studies used different GCMs. Nonetheless, our main results of decreasing globally averaged atmospheric water vapour content and surface temperature, and increasing day-night temperature contrasts, as land fraction increases are in agreement with theirs.

Other factors not accounted for here can have significant climate impacts. For instance, dust reduces temperatures on large, dry continents \citep{Boutle2020}. Colder planets could have hydrohalite, a type of salt crust precipitating out of very cold sea ice \citep{Shields2018}. Hydrohalite, which has a high infrared albedo, could cause a strong cooling feedback on a cold enough planet. The inclusion of a dynamic ocean should also affect the climate through increased redistribution of heat to the nightside, particularly in low-land-fraction SubCont models \citep{Hu2014, Salazar2020}.

On longer timescales, the stability of these climates will depend on silicate weathering, which requires rain falling on land (e.g., \citealt{Walker1981, pierrehumbert2010, Edson2012, Maher2014, Menou2015, Abbot2016, Paradise2019, Graham2020}). The weatherable surface area in SubCont models is limited by either the continent size or the spatial extent of the precipitation, since large continents are mostly dry. SubOcean models also have very little precipitation on land because rain mostly falls near the substellar point. RandCont models have more land precipitation, but still mostly near the substellar point, and it is scarce at high land fraction. We therefore expect weathering rates to be low in this general circulation regime. Seafloor weathering \citep{Krissansen-Totton2018, Hayworth2020} may increase climate stability on planets with large oceans.

Varying only the land fraction and configuration, we have generated differences of up to 20~K in globally averaged surface temperature, and several orders of magnitude in globally averaged atmospheric water vapour. Land albedo variations will introduce further temperature uncertainty. The temperature differences caused by pCO$_2$ and instellation changes (figures \ref{fig:SCtrends} and \ref{fig:SOtrends}) exceed those related to land fraction and configuration in our simulations, suggesting that atmospheric composition is a dominant source of uncertainty in M-Earth climates. Nevertheless, even if the atmosphere's composition is measured, land introduces a sizeable additional climate uncertainty.

\acknowledgments{
EM is supported by a Natural Science \& Engineering Research Council Post-Graduate Scholarship and by the University of Toronto Faculty of Arts and Science. AP is supported by an Ontario Graduate Scholarship.

The University of Toronto, where most of this work was performed, is situated on the traditional land of the Huron-Wendat, the Seneca, and the Mississaugas of the Credit. Some of this work was also performed on the traditional land of the Haudenosaunee and Anishinabeg nations in what is now known as Montreal, and on the traditional land of the Mi’kmaq in present-day Nova Scotia. We also acknowledge that this work required the extensive use of electricity-intensive supercomputer time; both hydroelectricity production and extraction of materials required to build computers have a substantial environmental impact, often on Indigenous land.
}

\bibliography{references}{}
\bibliographystyle{aasjournal}

\end{document}